# Photoionization profiles of metal clusters and the Fowler formula

Abhinav Prem and Vitaly V. Kresin

*Department of Physics and Astronomy, University of Southern California,
Los Angeles, California 90089-0484*

**Abstract**

Metal cluster ionization potentials are important characteristics of these "artificial atoms," but extracting these quantities from cluster photoabsorption spectra, especially in the presence of thermal smearing, remains a big challenge. Here we demonstrate that the classic Fowler theory of surface photoemission does an excellent job of fitting the photoabsorption profile shapes of neutral In$_{n=3-34}$ clusters [Wucher et al., New J. Phys. **10**, 103007 (2008)]. The deduced ionization potentials extrapolate precisely to the bulk work function, and the internal cluster temperatures are in close agreement with values expected for an ensemble of freely evaporating clusters. Supplementing an earlier application to potassium clusters, these results suggest that the Fowler formalism, which is straightforward and physical, may be of significant utility in metal cluster spectroscopy. It is hoped also that the results will encourage a comprehensive theoretical analysis of the applicability of bulk-derived models to cluster photoionization behavior, and of the transition from atomic and molecular-type to surface-type photoemission.

Photoionization measurements are a common probe of the structure of nanoclusters, and appearance energies (ionization potentials) $E_I$ serve as one of the primary characteristics of cluster properties. The dependence of $E_I$ on the number of atoms in a metal cluster, $n$, continues to attract interest [1]. It is often a challenge, however, to extract accurate $E_I$ from experimental ionization yield curves. Indeed, signals at threshold can become weak and hard to extrapolate; in addition, clusters in beams are frequently vibrationally excited and the "thermal tail" smears out the threshold region.

As a consequence, one has to resort to threshold-fit models or assumptions. Since a unified theory doesn't (yet) exist, a variety of schemes have been employed: linear, power-law or exponential extrapolations, error-function fits, thermal-oscillator models, etc. (see, e.g., [2-5] and references therein).

For bulk metal surfaces, on the other hand, there exists a long-standing, well defined expression for the near-threshold photoelectron yield function [6]. The derivation, due to Fowler [7], evaluates the flux of those conduction electrons whose kinetic energy of motion perpendicular to the surface, plus the energy $h\nu$ contributed by an absorbed photon, exceed the work function $\phi_0$. At finite temperatures, when thermal smearing of the Fermi–Dirac distribution is accounted for, the photoelectron yield has the form

$$\log\left(\frac{Y}{T^2}\right) = B + \log f\left(\frac{h\nu - \phi_0}{k_B T}\right). \qquad (1)$$

Here $B$ is a coefficient incorporating constants and instrumental parameters, and $f$ is an integral over the distribution function, expressible via a series expansion [7,8]:

$$f(x) = \begin{cases} e^x - e^{2x}/4 + e^{3x}/9 - \ldots & (x \leq 0); \\ \pi^2/6 + x^2/2 - \left(e^{-x} - e^{-2x}/4 + e^{-3x}/9 - \ldots\right) & (x \geq 0). \end{cases} \qquad (2)$$

A plot of $\log(Y/T^2)$ vs. $(h\nu-\phi_0)/k_B T$ is known as a ''Fowler plot,'' and by fitting data to the universal curve $\log f(x)$ one obtains the work function.

The same method has been found to apply successfully to coinage metal aerosols [9] and to alkali-metal nanoparticles in beams [10]. Several years ago, to probe whether smaller clusters may be amenable to the same approach, we showed [11] that even for potassium clusters of 30-100 atoms [12] the Fowler fit yielded very good results for their appearance energies and internal temperatures (see Fig. 1).

This was surprising, because the small cluster realm pushes the formalism beyond its original domain: instead of a continuous spectrum of electron waves in motion towards a flat surface, the picture is rather that of a discrete standing wave spectrum enclosed within a finite curved boundary. On the other hand, there are some similarities as well: firstly, in both cases the electron system is in contact with a thermal bath (for clusters it is the bath of vibrations characterized by their microcanonical temperature [13]) and secondly, for a departing photoelectron the curvature of the surface may not be salient at short separations. Fuller theoretical guidance would be valuable [14-16].

Additional tests are needed to ascertain that the successful Fowler analysis of alkali-metal clusters was not an isolated coincidence. A test requires well mapped-out photoion yield curves of size-selected free clusters with reasonably well-defined internal temperatures [17]. This can be realized in a beam produced in a thermalization nozzle or via an evaporative ensemble cascade (see below).

Recently Wucher et al. [18] presented photoionization efficiency curves for $In_{n=3-34}$ clusters produced by sputtering and ionized by single-photon absorption from a tunable laser. The authors modeled the threshold behavior via a procedure based on Refs. [19,20] which convolutes an assumed linear post-threshold photoelectron production curve with an internal energy distribution function derived from a thermal population of polyatomic oscillators. To generate reasonable values it was found necessary to make ad hoc fits to the number of vibrational degrees of freedom contributing to electron emission.

The data in Ref. [18] provide a convenient test set for the Fowler plot method. This method is coherent, straightforward and efficient, but will it work with these cluster spectra?

The ionization curves for sputtered clusters were digitized, and the variables $E_I$ (in place of $\phi_0$), $T$ and $B$ were optimized for best overlap of the data curve with the universal function $f$ in Eq. (1) when plotted in the reduced coordinates $\log(Y/T^2)$ as the ordinate and $(h\nu - E_I)/k_BT$ as the abscissa [21,22]. Fig 2 shows examples of threshold curves and fits.

The obtained ionization potentials are listed in Table 1 and plotted in Fig. 3 together with the values deduced in Ref. [18], and earlier experimental estimates [23] for clusters from a cold laser vaporization source. At $n \approx 10\text{-}12$ the $E_I$ values begin to follow a $\sim 1/R$ decrease, where $R \propto n^{1/3}$ is the cluster radius. (One may speculate whether such scaling could be regarded as a criterion for the onset of metallicity [24,25], since it is related to the screening of external charge by cluster electrons, see below.)

Writing down the well-known scaling relation (see, e.g., the reviews [1,2,26])

$$E_I \approx \phi_0 + \alpha \frac{e^2}{R}, \tag{3}$$

and performing linear regression for $n=12\text{-}34$, it is gratifying to find that the ionization potentials obtained from the Fowler fit extrapolate precisely to the bulk work function of polycrystalline indium [27]. Additionally, if the electron density parameter of metallic indium, $r_s=2.41a_0$ [28], is used to relate size and radius via $R=r_s(3n)^{1/3}$, the coefficient in Eq. (3) is found to be $\alpha \approx 0.46$. This is not far from the value for potassium clusters (Fig. 1) and from the frequently cited $\alpha \approx 3/8$ based on a semiclassical amalgamation of the work function and the image charge potential [26].

Note that the appearance potentials deduced by the thermal-oscillator fit [18] (the other set of points in Fig. 3) have a lower slope $\alpha$ and extrapolate to a much higher $\phi_0$ value.

Table 1 also lists internal cluster temperatures derived from the Fowler fits. Compared to the model in Ref. [18], the present values are lower by an average of ~60 K for $N<25$, and ~290 K for $N \geq 25$. The magnitude of these temperatures is very reasonable. Indeed, according to evaporative ensemble theory (see, e.g., Ref. [29] and references therein), vibrationally excited clusters produced in a hot source undergo prompt evaporation cascades *en route* to the detector, and reach an average internal temperature $k_BT \approx D/G$, where $D$ is the cluster dissociation energy and $G$ is the "Gspann parameter" [30]. The latter equals $\ln(t/\tau)$, where $t$ is the beam flight time and $\tau$ is the characteristic evaporation time, and has a typical magnitude of $G \sim 25\text{-}30$ [31].

Using the bulk heat of sublimation of indium [27] as an estimate, $D \approx 2.5$ eV, we find an expected temperature of ~1000-1200 K, in very good agreement with the range of values in Table 1 (the average of all the values listed is $\approx 1100$ K [32]). The aforementioned Fowler-type analysis of $K_n$ clusters [11] likewise produced cluster temperatures precisely in the expected range.

Thus, an application to a set of In cluster photoionization efficiency data has reaffirmed the robustness of the Fowler surface photoemission treatment. This is a tractable and physical method, and is now seen to provide excellent fits to ionization profile shapes for metal clusters of different sizes and materials. The fits yield directly the ionization potentials and the internal cluster temperatures; and both appear accurate, manifesting appropriate magnitudes and scaling behavior.

Since the formalism is efficient as well as accurate, it may become productive in cluster ionization spectroscopy. At the same time, it is interesting and important to explore theoretically the basis of its applicability to size-quantized cluster systems, as well as the general subject of the transition from atomic and molecular photoionization profiles to those of bulk surface photoemission.

We appreciate C. Stark's contribution at the initial stage of this project, and support by the NSF and the USC Undergraduate Research Associates Program.

**Table 1.** Appearance energies and temperatures (rounded to the nearest 50 K) of In$_n$ clusters, determined by a Fowler fit to the photoionization data of Ref. [18].

| $n$ | $E_I$ (eV) | $T$ (K) |
|---|---|---|
| 3 | 5.37 | 2250 |
| 4 | 5.3 | 1100 |
| 5 | 5.52 | 1250 |
| 6 | 5.63 | 1200 |
| 7 | 5.67 | 1300 |
| 8 | 5.6 | 1100 |
| 9 | 5.54 | 1000 |
| 10 | 5.59 | 1100 |
| 11 | 5.56 | 1150 |
| 12 | 5.56 | 1050 |
| 13 | 5.6 | 1250 |
| 14 | 5.59 | 1300 |
| 15 | 5.45 | 1200 |
| 16 | 5.51 | 1000 |
| 17 | 5.48 | 1200 |
| 18 | 5.42 | 1000 |
| 19 | 5.4 | 1100 |
| 20 | 5.37 | 950 |
| 21 | 5.37 | 1100 |
| 22 | 5.38 | 1100 |
| 23 | 5.31 | 1050 |
| 24 | 5.3 | 1000 |
| 25 | 5.28 | 1000 |
| 26 | 5.32 | 1150 |
| 27 | 5.36 | 1200 |
| 28 | 5.01 | 600 |
| 29 | 5.24 | 1300 |
| 30 | 5.24 | 1150 |
| 31 | 5.23 | 1250 |
| 32 | 5.13 | 1100 |
| 33 | 5.18 | 1300 |
| 34 | 5.14 | 1100 |

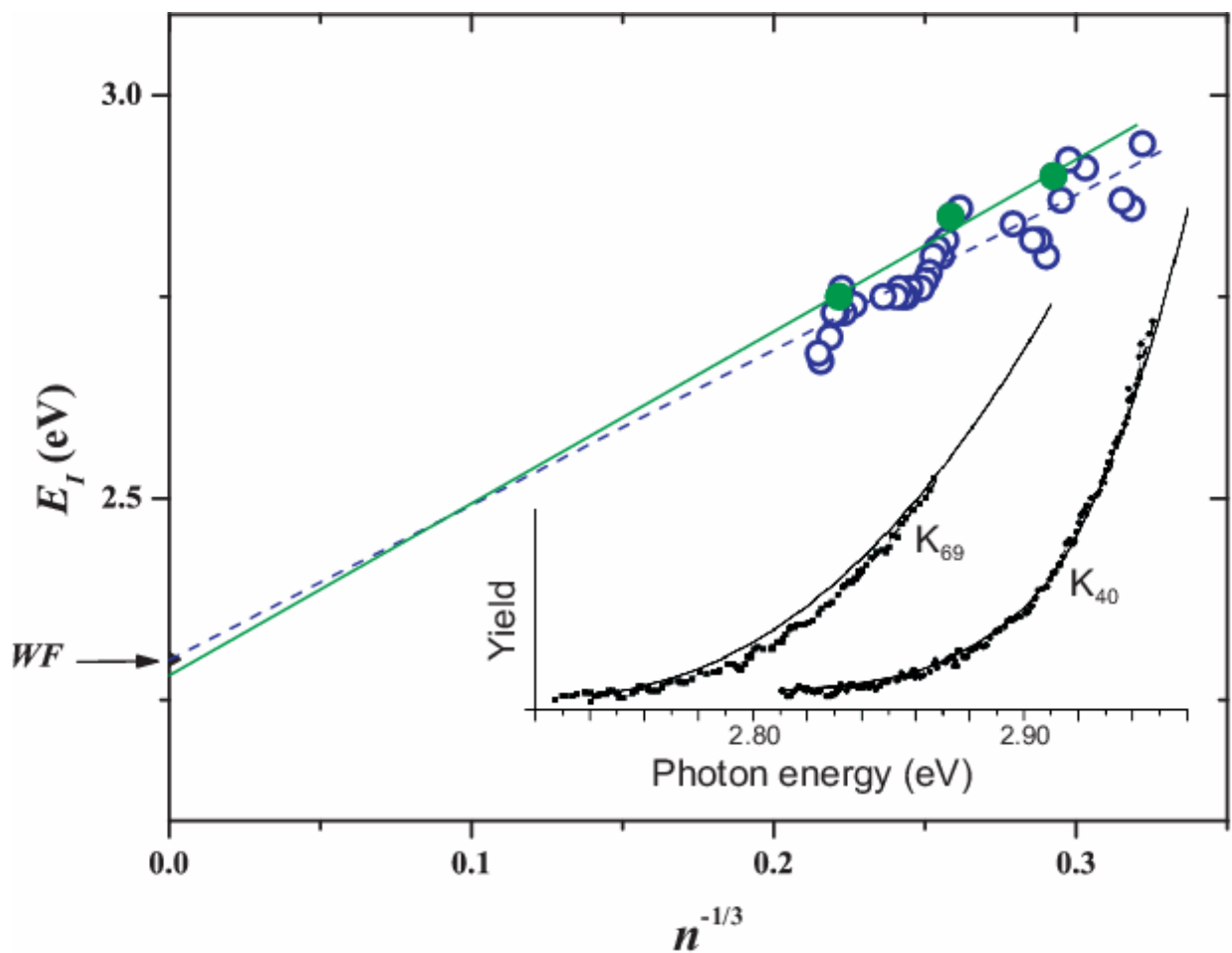

**Fig. 1.** Appearance energies of $K_{30 \leq n \leq 101}$ clusters determined by a Fowler fit [11] to photoionization profiles [12]. With $R=r_s n^{1/3}$ ($r_s$=4.86$a_0$), linear regression gives $E_I$ = 2.30 eV +0.34$e^2/R$ for all points, or 2.29 eV+0.38$e^2/R$ for the spherical closed-shell clusters. The potassium metal work function is 2.3 eV. Insert: examples of threshold data [12] and finite-temperature Fowler functions (solid lines) [11].

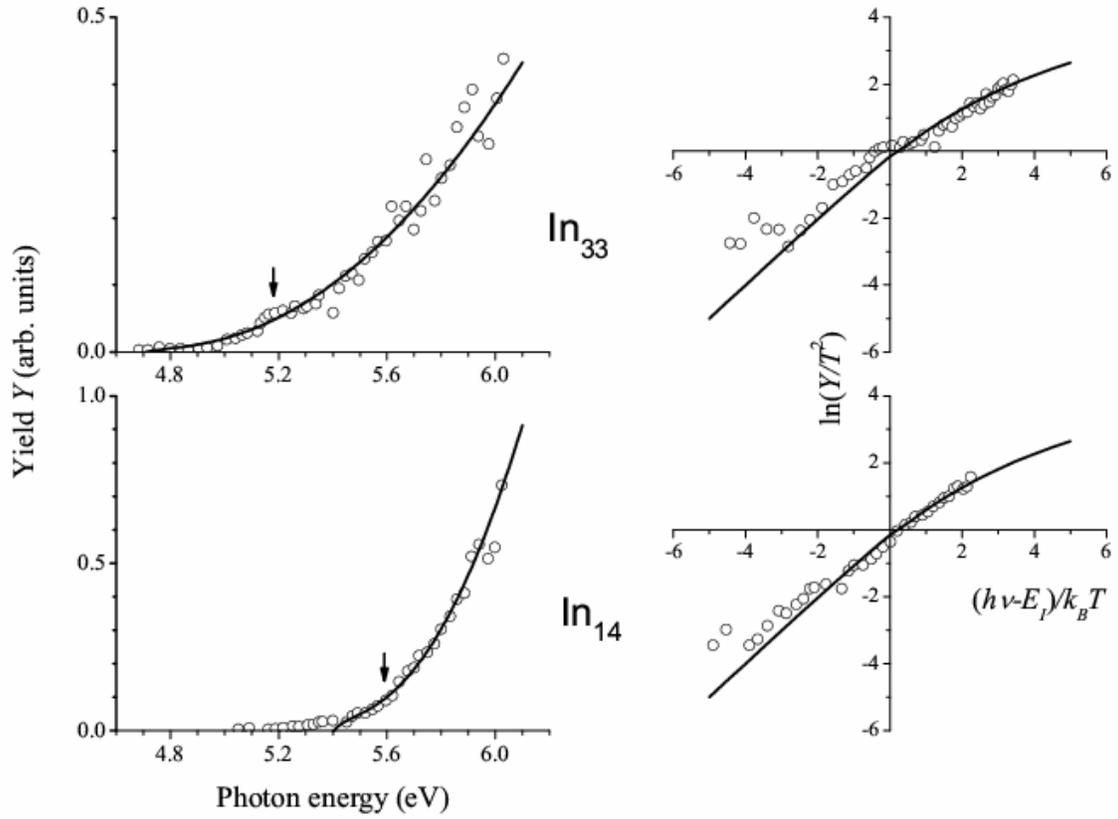

**Fig. 2.** Examples of Fowler fits for indium clusters (top: $In_{33}$, bottom: $In_{14}$). On the right are Fowler plots following Eq. (1) (lines: finite-temperature Fowler functions, dots: experimental data [18]). On the left are the corresponding ionization threshold profiles (line: Fowler functions, dots: experimental data, arrows: deduced appearance energies $E_I$).

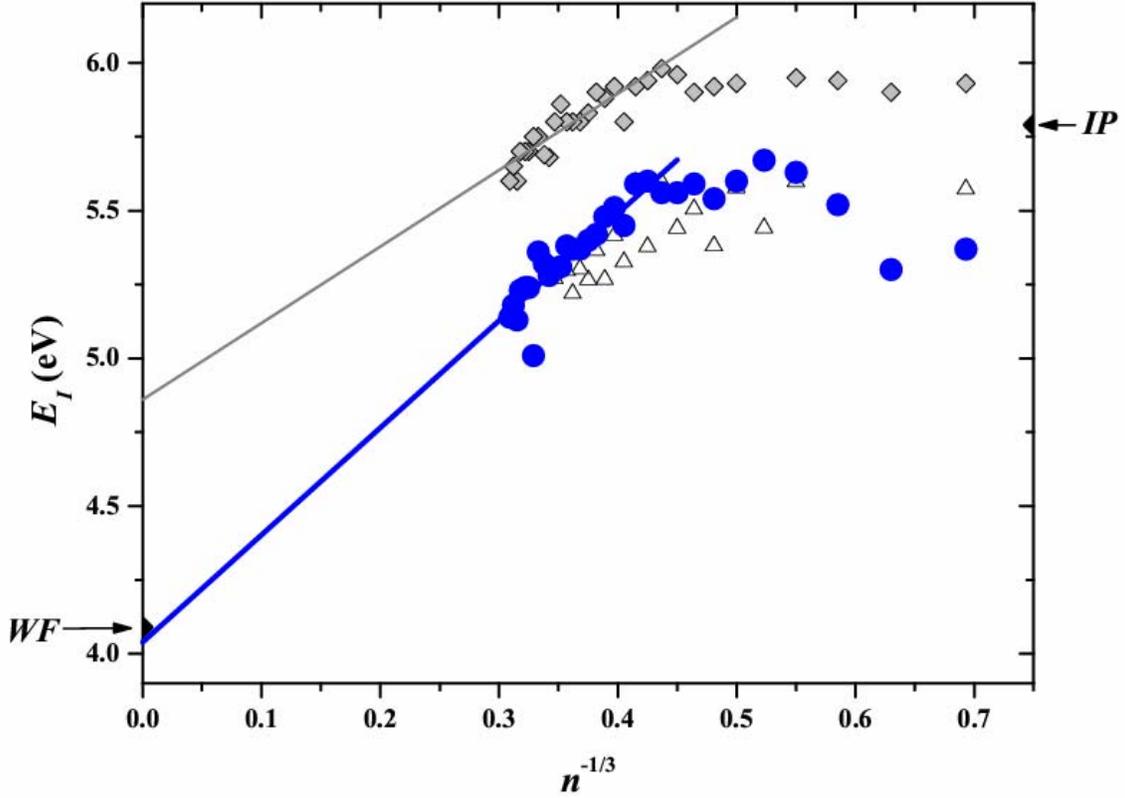

**Fig. 3.** (Color online) Blue solid circles: appearance energies of $In_{3 \leq n \leq 34}$ clusters determined by a Fowler fit to photoionization profiles [18], as described in the text and illustrated in Fig. 2. With $R=r_s(3n)^{1/3}$ ($r_s=2.41a_0$), linear regression for $n>12$ (heavy blue line) gives $E_I$ = 4.04 eV + $0.46e^2/R$. The $y$-intercept is in excellent agreement with the polycrystalline work function of indium, 4.09 eV. The grey diamonds are $E_I$ values deduced from the same data via a thermal-oscillator fit in Ref. [18]. The triangles are $E_I$ values from Ref. [23] estimated from measurements using a cold laser-vaporization cluster source. For reference, the atomic ionization potential is marked on the right-hand axis.